\documentclass[sigconf, 10pt]{acmart}

\usepackage{fancyhdr}
\usepackage{graphicx}
\usepackage{bm} 
\usepackage{subcaption} 
\usepackage{soul}
\usepackage{amsmath}

\setcopyright{acmlicensed}
\copyrightyear{2024}
\acmYear{2024}
\acmDOI{10.1145/3636534.3697324} 

\acmConference[WiNTECH '24]{The 18th ACM Workshop on Wireless Network Testbeds, Experimental evaluation and Characterization}{November 18,
  2024}{Washington, D.C., USA}
  
\begin{document}
\copyrightyear{2024}
\acmYear{2024}
\setcopyright{acmlicensed}\acmConference[ACM MobiCom '24]{The 30th Annual International Conference on Mobile Computing and Networking}{November 18--22, 2024}{Washington D.C., DC, USA}
\acmBooktitle{The 30th Annual International Conference on Mobile Computing and Networking (ACM MobiCom '24), November 18--22, 2024, Washington D.C., DC, USA}
\acmDOI{10.1145/3636534.3697324}
\acmISBN{979-8-4007-0489-5/24/11}

\title{Experimental Validation of a 3GPP Compliant 5G-Based Positioning System}


\author{Sarik Dhungel}
\affiliation{
\institution{NC State University}
\city{Raleigh, NC}
\country{USA}
}
\email{sdhunge@ncsu.edu}
\authornotemark[1]

\author{Gaurav Duggal}
\affiliation{
\institution{Virginia Tech}
\city{Blacksburg, VA}
\country{USA}}
\email{gduggal@vt.edu}
\authornote{Both authors contributed equally to this research.}


\author{Dara Ron}
\affiliation{
\institution{NC State University}
\city{Raleigh, NC}
\country{USA}
}
\email{dron@ncsu.edu}

\author{Nishith Tripathi}
\affiliation{
\institution{Virginia Tech}
\city{Blacksburg, VA}
\country{USA}}
\email{nishith@vt.edu}

\author{R. Michael Buehrer}
\affiliation{
\institution{Virginia Tech}
\city{Blacksburg, VA}
\country{USA}}
\email{rbuehrer@vt.edu}

\author{Jeffrey H. Reed}
\affiliation{
\institution{Virginia Tech}
\city{Blacksburg, VA}
\country{USA}}
\email{reedjh@vt.edu}

\author{Vijay K Shah}
\affiliation{
\institution{NC State University}
\city{Raleigh, NC}
\country{USA}}
\email{vijay.shah@ncsu.edu}


\renewcommand{\shortauthors}{Dhungel et al.}

\begin{abstract}
The advent of 5G positioning techniques by 3GPP has unlocked possibilities for applications in public safety, vehicular systems, and location-based services. However, these applications demand accurate and reliable  positioning performance, which has led to the proposal of newer positioning techniques. To further advance the research on these techniques, in this paper, we develop a 3GPP-compliant 5G positioning testbed, incorporating gNodeBs (gNBs) and User Equipment (UE). The testbed uses New Radio (NR) Positioning Reference Signals (PRS) transmitted by the gNB to generate Time of Arrival (TOA) estimates at the UE. We mathematically model the inter-gNB and UE-gNB time offsets affecting the TOA estimates and examine their impact on positioning performance.
Additionally, we propose a calibration method for estimating these time offsets. Furthermore,  we investigate the environmental impact on the TOA estimates. Our findings are based on our mathematical model and supported by experimental results. 
\end{abstract}

\ccsdesc[500]{Networks~Network experimentation}
\ccsdesc[500]{Hardware~Hardware validation}

\keywords{5G Positioning, O-RAN, 3GPP, TOA, TDOA, and OAI5G.}

\maketitle

\section{Introduction}


With the rapid increase in Internet of Things (IoT) devices supporting fifth-generation (5G) cellular networks over the recent years, interest in 5G positioning has also grown simultaneously because of its potential for the seamless integration of precise positioning services along with communication. 5G cellular-based positioning has the potential to offer positioning for diverse applications since cell phones are ubiquitous. Some applications are for public safety \cite{albaneseetal, Dureppagari_publicsafety}, vehicular positioning \cite{ge2024v2x}, healthcare \cite{witrisal2016high}, and asset tracking \cite{jiang2021survey}.
With its diverse applications, 5G positioning presents unique challenges. For example, in public safety scenarios such as firefighting, there is a need to improve the vertical axis positioning accuracy \cite{firstnetroadmap}. Vehicular positioning performance is adversely affected in urban scenarios. With the ever-increasing demand for positioning performance, we note that a major contributing factor is environmental effects such as the probability of the wireless link being Line-of-Sight (LOS) and various multipath effects \cite{Gifford2020}. 
Further, new techniques such as diffraction-based positioning \cite{duggal20243dpositioningusingnew} have been proposed and have shown notable improvement in theoretical positioning performance. Taking such techniques to the real world requires experimental validation of theoretical results. Although some studies have conducted experimental evaluations of 5G positioning testbeds \cite{mundlamuri2024, reviewer_18,exp_assessment_of_sdr_5g_pos}, more comprehensive validation of time synchronization models and investigations into hardware, software, and environment-related implementation challenges are still required. 



In our study, we address these gaps by providing a 3GPP-compliant 5G positioning testbed built using Software Defined Radios (SDRs) and open-source cellular software called OpenAirInterface (OAI) that has the potential to facilitate the accelerated testing and deployment of advanced positioning algorithms in real-world scenarios. Our proposed solution relies on the use of mathematical modeling, followed by experimental results. The main contributions are outlined below.  

$\bullet$ \textit{Development of a 3GPP compliant Positioning testbed:} Using various hardware and software components, we emulate 5G gNBs and UE to achieve 2D positioning capability based on extracting TOA measurements using the 5G downlink PRS.

$\bullet$ \textit{Impact of Multipath on positioning:} Based on outdoor and indoor experiments, we investigate the impact of multipath on TOA estimation which ultimately affects positioning performance. We highlight the role of selecting the appropriate signal bandwidth based on characterizing the signal propagation environment using metrics such as {\em delay spread}.

$\bullet$ \textit{Resolving Time synchronization errors:} Despite employing a hardware-based time synchronization procedure in our testbed, the gNBs cannot achieve perfect inter-gNB time synchronization. We mathematically model these residual inter-gNB timing errors, demonstrate their impact on the positioning, and then propose a calibration procedure to estimate these offsets thus making the testbed 3GPP compliant. Further experiments are conducted to validate the time stability and validity of the estimated timing offsets resulting from the calibration procedure.
section{5G D PRS Specifications}

\subsection{5G-Positioning Architecture}
\label{5g_enhanced_pos_est}
\begin{figure}[t]
  \centering
  \includegraphics[width=0.8\linewidth]{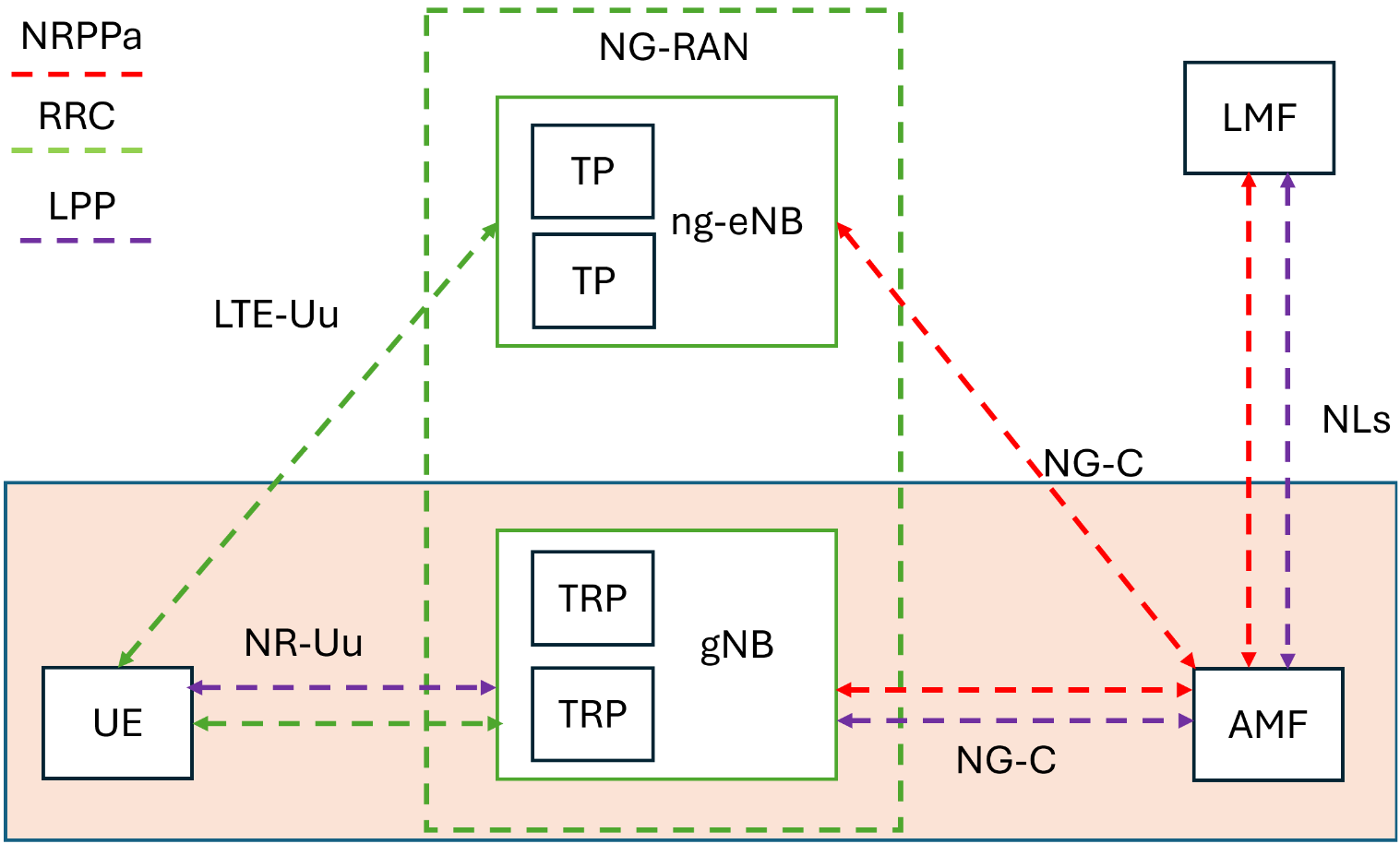}
  \captionsetup{font=small}
   \vspace{-0.1in}
  \caption{\footnotesize 5G Positioning architecture as per 3GPP Release 16 \cite{10041419}}
  \label{fig_5G_pos}
  \Description{5G Positioning architecture as per 3GPP Release 16}
\end{figure}

The system architecture (see Figure \ref{fig_5G_pos}) consists of 5G UE,  Next Generation (NG) Radio Access Network (RAN), and the 5G Core Network (CN). NG RAN consists of the evolved LTE eNodeB (eNB) that can communicate with a 5G CN and fully 5G gNB.
5G CN comprises the Access Mobility Function (AMF) and Location Management Function (LMF). AMF maintains continuous communication and location updates by managing the connection and mobility of the 5G UE. LMF receives necessary location service requests, including measurements and information about the UE from NG-RAN via AMF for positioning \cite{whitepaper}. Commercial 5G networks typically use 5G UEs that communicate with the 5G gNBs using the 5G NR air interface (NR-Uu). Our study uses 5G gNB to transmit a downlink NR PRS signal over NR-Uu to the UE for location estimation. This 5G positioning architecture supports methods that utilize Time Difference of Arrival (TDOA) to determine the location of the UE.  

\subsection{PRS Frame Structure and Configuration}

Our positioning system focuses on utilizing the DL PRS signal transmitted from three gNBs to a single UE for TOA measurements at the UE. PRS is a wideband signal with staggered resource element (RE) patterns, offering excellent auto-correlation and low cross-correlation properties. These features make them particularly well-suited for accurate positioning applications. TS 38.211, in \cite{3GPPTS38211} provides details on the methods for generating PRS reference-signal sequences.

PRS can be flexibly configured to map onto the 5G frame structure with granularity down to the symbol level. This flexibility allows PRS to be mapped in the time and frequency domains according to deployment requirements. Higher numerology with larger subcarrier spacing offers enhanced timing resolution and reduced latency, making it ideal for precise positioning in urban environments. The length of each 5G frame is $10$ ms in duration and is divided into $10$ sub-frames. Depending on the numerology used, each sub-frame can contain one or multiple slots. In our case, we employ numerology $1$ with a $30$ kHz sub-carrier spacing (scs) and $106$ Physical Resource Blocks (PRBs), with each PRB comprising $12$ sub-carriers in the frequency domain and $14$ symbols per $0.5$ms slot, yielding a total bandwidth of $106\times12\times 30$ KHz $= 38.16$ MHz.

Each gNB can be configured with a DL PRS configuration for signal transmission. The TOA measurement is estimated based on the PRS signals received by the UE. The PRS configuration can be described as follows- ($1$) \textit{PRS Resource Set Period} refers to the periodicity of the slots; ($2$) \textit{PRS Resource Offset} defines the slot offset of each PRS resource; ($3$) \textit{PRS Resource Repetition} indicates the repetition factor or the number of slots allocated for each PRS resource within a single PRS Resource Set Period; and ($4$) \textit{PRS Resource Time Gap} refers to the slot offset between two consecutive PRS repetitions.

\begin{figure}[t]
 \centering
      \includegraphics[width = 1\linewidth]{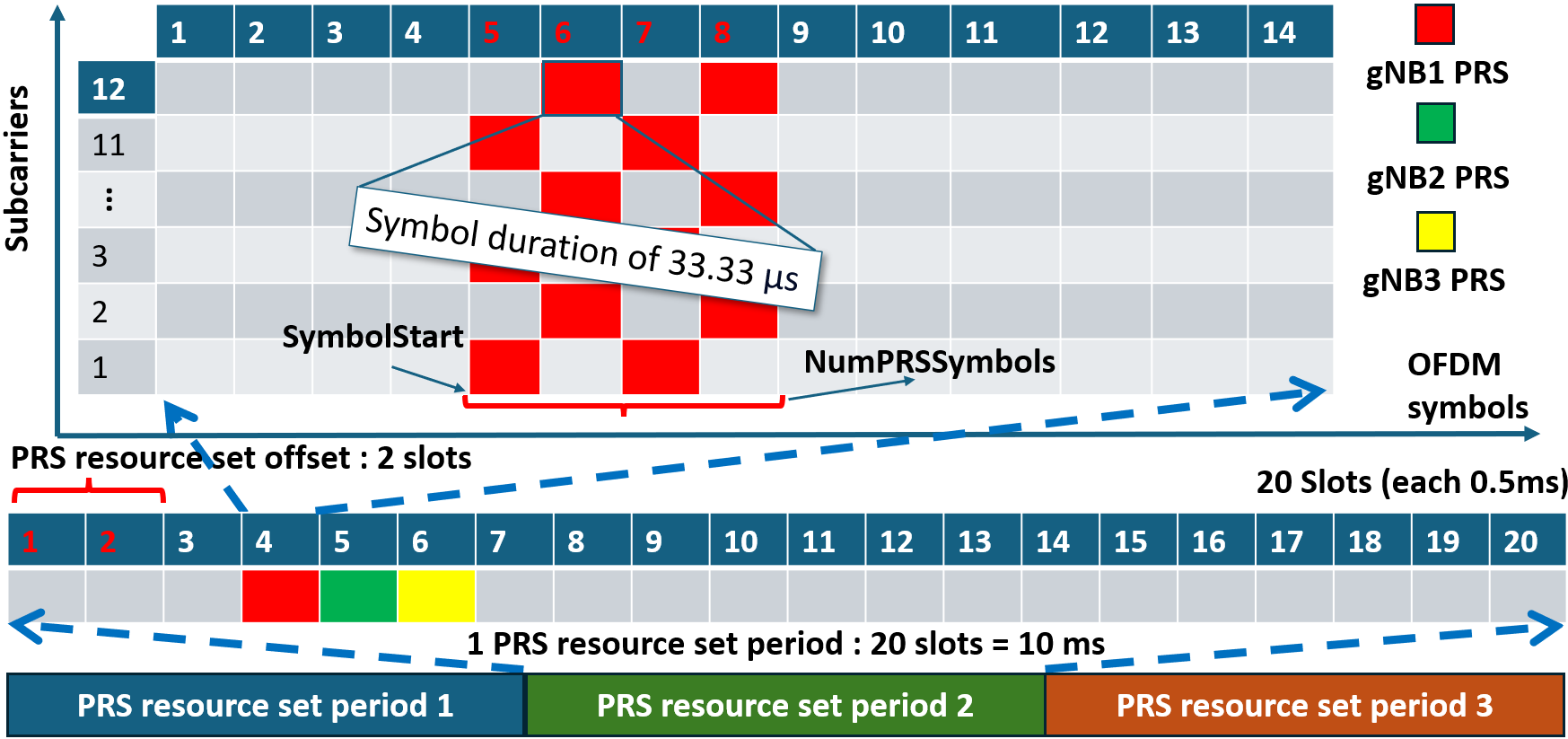}
      \captionsetup{font=small}
          \caption{PRS configurations for the UE and three gNBs to transceive 5G positioning PRS signals.}
          \Description{PRS configurations for the UE and three gNBs to transceive 5G positioning PRS signals.}
          \label{comb2_4symb}
 \end{figure}

Figure \ref{comb2_4symb} provides a diagrammatic representation of the PRS configuration, which includes a \textit{PRS Resource Set Period} of $20$ slots and \textit{PRS Resource Set Offset} of $2$ slots, a \textit{PRS Resource Offset} of $1$, $2$, and $3$ slots for gNB1, gNB2, and gNB3 respectively, a \textit{PRS Resource Repetition} of $1$ for each PRS resource, and a \textit{PRS Resource Time Gap} of $1$ slot.

These configurations ensure that PRS from different gNBs are mapped to unique slots, thereby preventing signal overlap between them. The configuration described above focuses on the slot-level configuration. We now concentrate on the configuration in the frequency domain and symbols inside the slot. Within each time slot, the number of consecutive Orthogonal Frequency Division Multiplexing (OFDM) symbols is specifically allocated for each PRS resource. The PRS configuration parameters for this are: ($1$) \textit{Symbol Start}, ($2$) \textit{Number of OFDM Symbols}, ($3$) \textit{RB Offset}, ($4$) \textit{Number of RBs}, and ($5$) \textit{Comb Size}. Detailed information about these parameters can be found in \cite{3GPPTS38211}.

\section{5G Positioning Testbed Overview}
\label{section_positioning_testbed_overview}
This section describes the prototyping of our 5G positioning testbed using SDRs and the OAI cellular software stack. Using this setup as a reference, we investigate the challenges of accurately estimating the UE's 2D location and explore initial mitigation techniques to address these issues.

\begin{figure}[t]
\centering
     \includegraphics[width=0.9\linewidth]
     {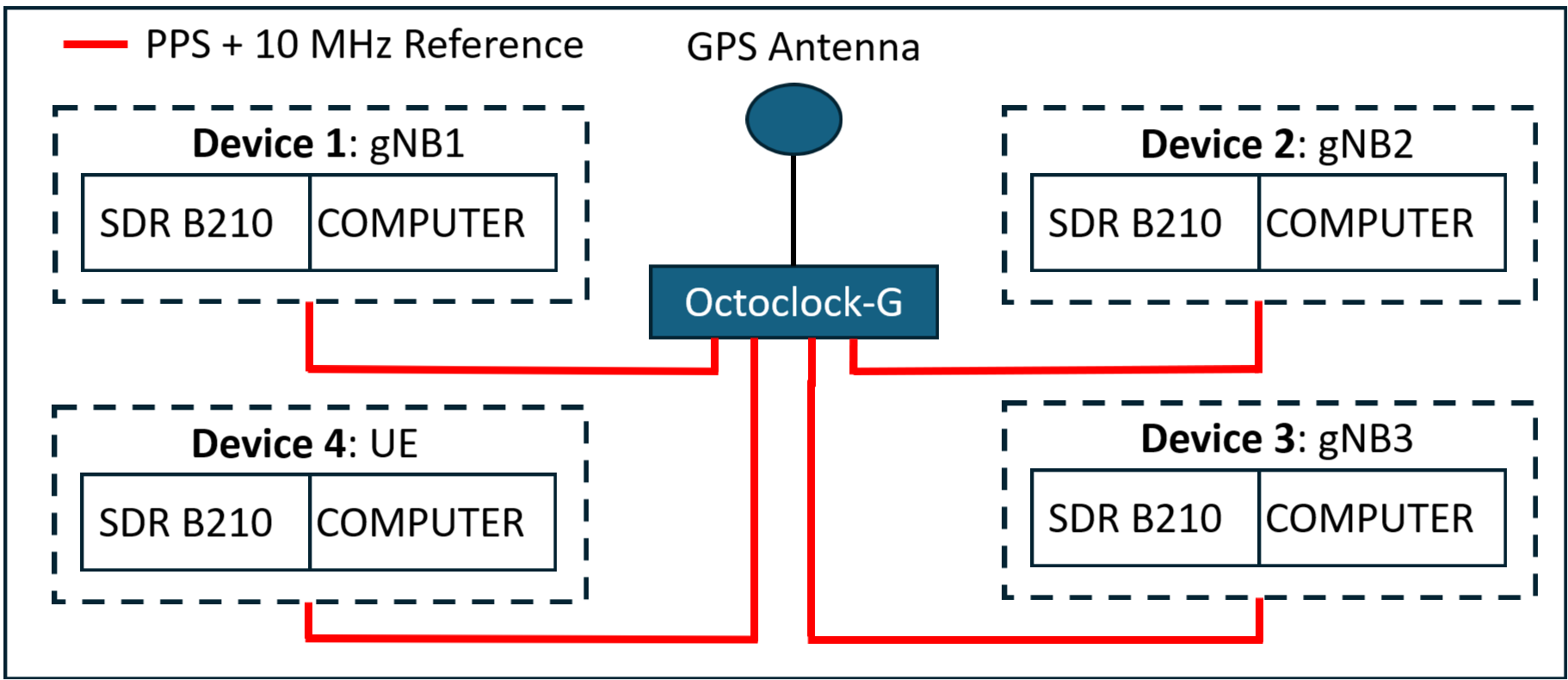}
         \captionsetup{font=small}
         \caption{Testbed setup: Three gNBs and one UE are connected to an Octoclock for time and frequency synchronization.}
         \label{fig_testbed}
         \Description{Positioning setup}
\end{figure}

Our positioning testbed consists of four devices as shown in Figure \ref{fig_testbed}. A device comprise an SDR B210 and a desktop computer. Each B210 is connected to the octoclock-G, used for both time and frequency synchronization. An Octoclock is equipped with GPS-disciplined oscillator for time synchronization between devices within 50 ns.
The frequency stability of the device oscillators achieved with the octoclock is 20 parts-per-billion (ppb) \cite{octoclock_manual}. In each device, the computer runs an open source 3GPP compliant 5G PHY layer software stack - OAI \cite{openairinterfaces} and each device can be configured to operate as a UE or a gNB. We configure one device as a UE, whereas the other three devices are configured as gNBs. We postpone further discussion about the testbed to section \ref{section_experimental_setup} after development of the mathematical model of the system.

\section{Mathematical Model of 5G Testbed }
\label{section_system_model}

\begin{figure}[t]
  \centering
  \includegraphics[width=0.7\linewidth]{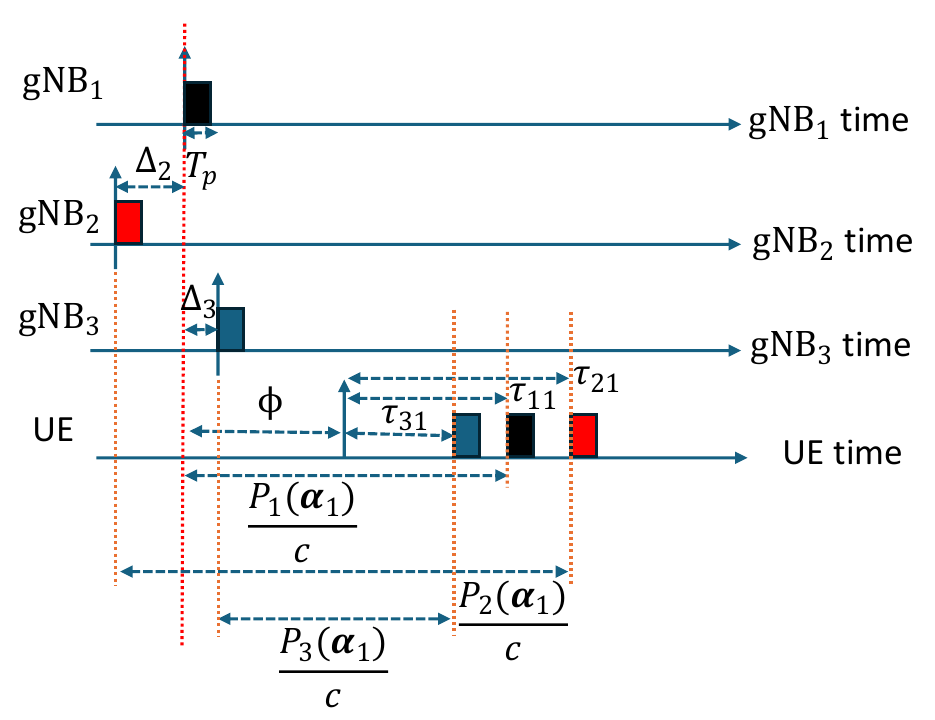}
  \captionsetup{font=small}
  \vspace{-0.1in}
  \caption{\footnotesize System Model with three gNBs and one UE with unknown time offsets between each other. $\tau_{i,j}$ can be positive or negative depending upon the values of $\phi$ and $\Delta_{j}$.}
  \label{fig_offsets}
  \Description{System Model with three gNBs and one UE with unknown time offsets between each other.}
  \vspace{-0.01in}
\end{figure}

In the 3GPP 5G standards, the UE is assumed to have an unknown time offset relative to the gNBs, while all gNBs are considered perfectly synchronized. However, in our 5G testbed, despite hardware synchronization, residual unknown time offsets exist among devices, as noted in the octoclock specifications \cite{octoclock_manual}, illustrated in Figure \ref{fig_offsets}. In this section, we develop a mathematical model for these offsets and demonstrate how TOA information is extracted from channel estimates. We then describe a calibration process to achieve synchronized gNBs, while the UE maintains an unknown fixed time offset that can be jointly estimated with its position using standard TDOA techniques \cite{zekavat2019handbook}.

We assume that the UE is located at $\bm{\alpha}_i=[x_{n_i},y_{n_i}]^T$ and the $j^{th}$ gNB is located at $[x_{a_j},y_{a_j}]^T$. There are three gNBs in total i.e. $j \in \{1,2,3\}$. 
Therefore, the Euclidean distance $p_{j}(\bm{\alpha}_i)$ between the $j^{th}$ gNB and the $i^{th}$ UE location can be expressed as
{\small
\begin{equation}
    p_{j}(\bm{\bm{\alpha}}_i) = \sqrt{(x_{a_j}-x_{n_i})^2 + (y_{a_j}-y_{n_i})^2}.
\end{equation}
}
The signal propagation time along the first arriving path for the $j^{th}$ gNB is $\frac{p_{j}(\bm{\alpha}_i)}{c}$, where $c$ is the speed of light. 
Now, TOA measurements conducted at the UE can be expressed as
{\small
\begin{equation}
    \label{eq_TOA}
    \tau_{i,j} =  \frac{p_{j}(\bm{\alpha}_i)}{c} + \phi + \Delta_j + n_{i,j}\\
\end{equation}
}
Here, $\tau_{i,j}$ is the TOA estimate between the $j^{th}$ gNB and the $i^{th}$ UE position, 
$\phi$ is the gNB-UE time offset. Assuming $j=1$ as the reference gNB, we have the inter-gNB time offsets $\Delta_2$ and $\Delta_3$ as the unknown time offsets with respect to the reference gNB. Note, we have $\Delta_1=0$ by definition. The measurement noise at the UE from the $j^{th}$ gNB is $n_j$ and is considered to be zero mean Gaussian noise and independent between the different gNBs.

\subsection{Time Of Arrival Estimation}
\label{section_TOA_estimation_from_channel}

\begin{figure}[t]
  \centering
  \includegraphics[width=0.7\linewidth]{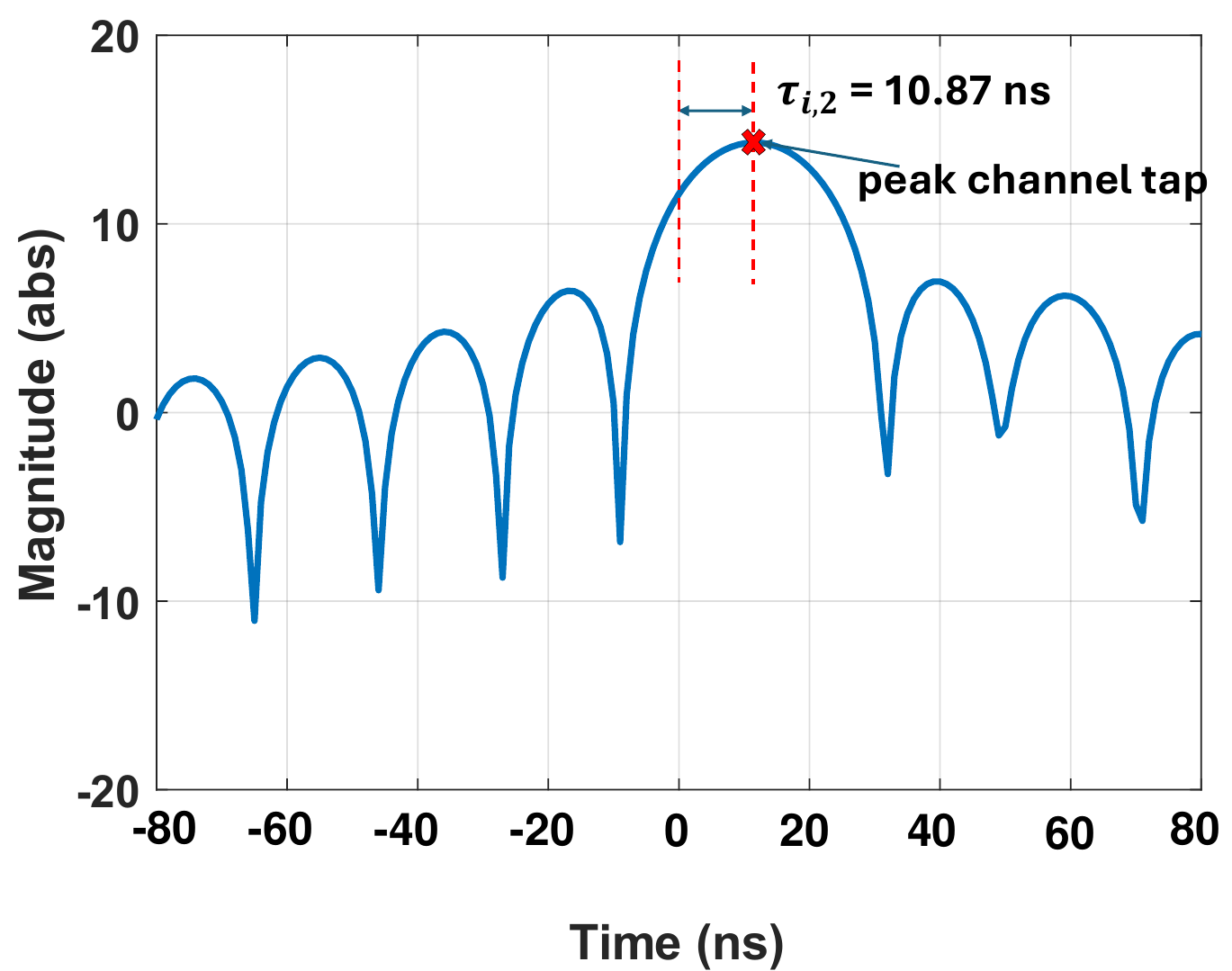}
  \captionsetup{font=small}
  \vspace{-0.1in}
  \caption{\footnotesize TOA ($\tau_{i,2}$) is the location of the `peak channel tap' obtained from the downlink channel estimate from gNB $2$. 
  }
  \label{fig_TOA_definition}
  \Description{TOA definition}
 \end{figure}

We define the first arriving path as the signal path between the gNB and UE with the shortest propagation time. In LOS scenarios, this is also the direct path. TOA is defined as the time at which the UE receives the gNB signal via the first arriving path (see Figure \ref{fig_TOA_definition}). Our objective is to estimate the TOA from each gNB using its downlink channel estimate. Each gNB has a unique PRS configuration, ensuring that its PRS signal is orthogonal to others and received interference-free at the UE. Thus, during each PRS resource set interval, we obtain three channel estimates, one from each gNB.

Assuming a multipath free environment, for each gNB, we expect to see a single peak in the associated downlink time domain channel estimate. This peak is associated with the PRS signal propagating from the gNB to the UE. Since we assumed LOS conditions, the TOA of the PRS signal corresponds to the signal propagation along the Euclidean distance between the gNB and UE. Now, in a given gNB channel estimate, the `peak channel tap' is defined to be the particular channel tap that corresponds to the maximum value. This peak channel tap is converted to the TOA (in seconds) by multiplying it with the channel tap to TOA conversion factor. 
For the B210 radios, the ADC sampling rate is $46.08$ MHz corresponding to a conversion factor $2.17\times  10^{-8}s/$channel tap. This means if the peak shifts by one channel tap, the TOA estimate shifts by $21.7$ ns. {\em OAI} converts the time domain channel to frequency domain, zero pads the frequency domain channel and then converts the resulting zero-padded channel back into time domain. This operation is known as \textit{digital interpolation} and this increases the number of channel taps digitally by interpolation, hence improving the conversion factor. 

In our testbed, after interpolation, we end up with $16$ times the number of time domain taps for the channel, thus increasing the conversion factor to $\frac{6.5m}{16}\approx 0.4m$. Note, in a LOS scenario with minimal multipath, digital interpolation improves the TOA resolution. Consider, a multipath scenario, where the downlink time domain channel estimate contains multiple peaks. Digital interpolation in this case will not improve the TOA resolution. 
As the separation between peaks decreases, they begin to overlap, causing the peak channel tap to no longer correspond to the first arriving path. To enhance TOA resolution, it's essential to reduce the width of the peaks associated with each multipath component in the downlink time-domain channel. This minimizes peak overlap and improves TOA accuracy. Since peak width is inversely proportional to signal bandwidth, in environments with significant multipath, both bandwidth and signal-to-noise ratio (SNR) are critical factors that determine TOA estimation performance.

 

\subsection{Calibration Process for Inter-gNB Time Offset Estimation} \label{section_calibration}

The $\pm 50ns$ of time synchronization accuracy offered by the octoclock results in the same error in the TOA measurements. Therefore, $\Delta_2$, $\Delta_3$ and gNB-UE time offset $\phi$ need to be accounted for since not accounting for them can lead to a ranging error of $\pm 15m$ which is insufficient for achieving good positioning performance. Since the Octoclock offers frequency synchronization of 20 ppb, we expect the time offsets to not change with time over the duration that the testbed is in use. We investigate this in  section \ref{section_results_time_offsets}.

An intermediary goal for positioning is to obtain the signal propagation time $\frac{p_j(\bm{\alpha}_i)}{c}$ from each gNB to the UE using the TOA estimates $\tau_{i,j}$. In equation \eqref{eq_TOA}, $\bm{\alpha}_i=[x_{n_i},y_{n_i}]^T$, $\Delta_1$, $\Delta_2$ and $\phi$ are unknown. With three gNBs, we have three TOA estimates, resulting in only three equations and an inability to estimate the UE position. This section outlines a calibration procedure to determine the inter-gNB time offsets $\Delta_2, \Delta_3$ while $\phi$ can be eliminated using the TDOA method.

With the reference gNB as $j=1$, we subtract equation \eqref{eq_TOA} for $j^{th}$ gNB from equation \eqref{eq_TOA} for $j=1$. On rearranging the terms, we obtain
{\small
\begin{equation}
    \label{eq_RSTD_defn}
    \Delta_j =  \underbrace{(\tau_{i,j} - \tau_{i,1})}_{\text{Measured RSTD}} - \underbrace{\frac{P_j(\bm{\alpha}_i) - P_1(\bm{\alpha}_i)}{c}}_{\text{True RSTD}} +\; n_{i,j} - n_{i,1},\;\; j\ne1.
\end{equation}
}
We define the Measured Reference Signal Time Difference (RSTD) as the difference between the TOA from the $j^{th}$ gNB and the TOA from the reference gNB. Similarly, the True RSTD is defined as the difference in Time of Flight (TOF) from the $j^{th}$ gNB and the reference gNB. Thus, the inter-gNB offset for gNB $j$ can be written as the difference between the Measured RSTD and the True RSTD. 
\par
For the calibration procedure, the UE is placed at known positions $\bm{\alpha}_i$ and since the gNB position is known as well, the true RSTD is a fixed known quantity. Further, in equation \eqref{eq_RSTD_defn}, we assume the noise in the TOA measurements is zero mean Gaussian and independent between gNBs. Thus adding two noise terms results in another zero mean Gaussian random noise which can be averaged out over several measurements. The calibration process is expressed as an equation as,

\begin{equation}
    \label{eq_est_delta2}
    \hat{\Delta}_j = \frac{1}{K} \sum_{i=1}^{K} \Bigl( (\tau_{i,j} - \tau_{i,1}) - \frac{P_j(\bm{\alpha}_i) - P_1(\bm{\alpha}_i)}{c} \Bigr)
\end{equation}

To summarize, we obtain $K$ TOA measurements for the $j^{th}$ gNB by placing the UE at $K$ known locations. The time offset for each non-reference gNB is obtained by averaging the difference between the measured RSTD and the true RSTD for the $j^{th}$ gNB. Note, here $j \ne 1$ and $\Delta_1=0$ by definition because it is the gNB time offset with respect to itself -- the reference gNB.

\subsection{TDOA Position Determination}
\label{section_tdoa_positioning}
\begin{figure}[t]
  \centering
  \includegraphics[width=0.7\linewidth]{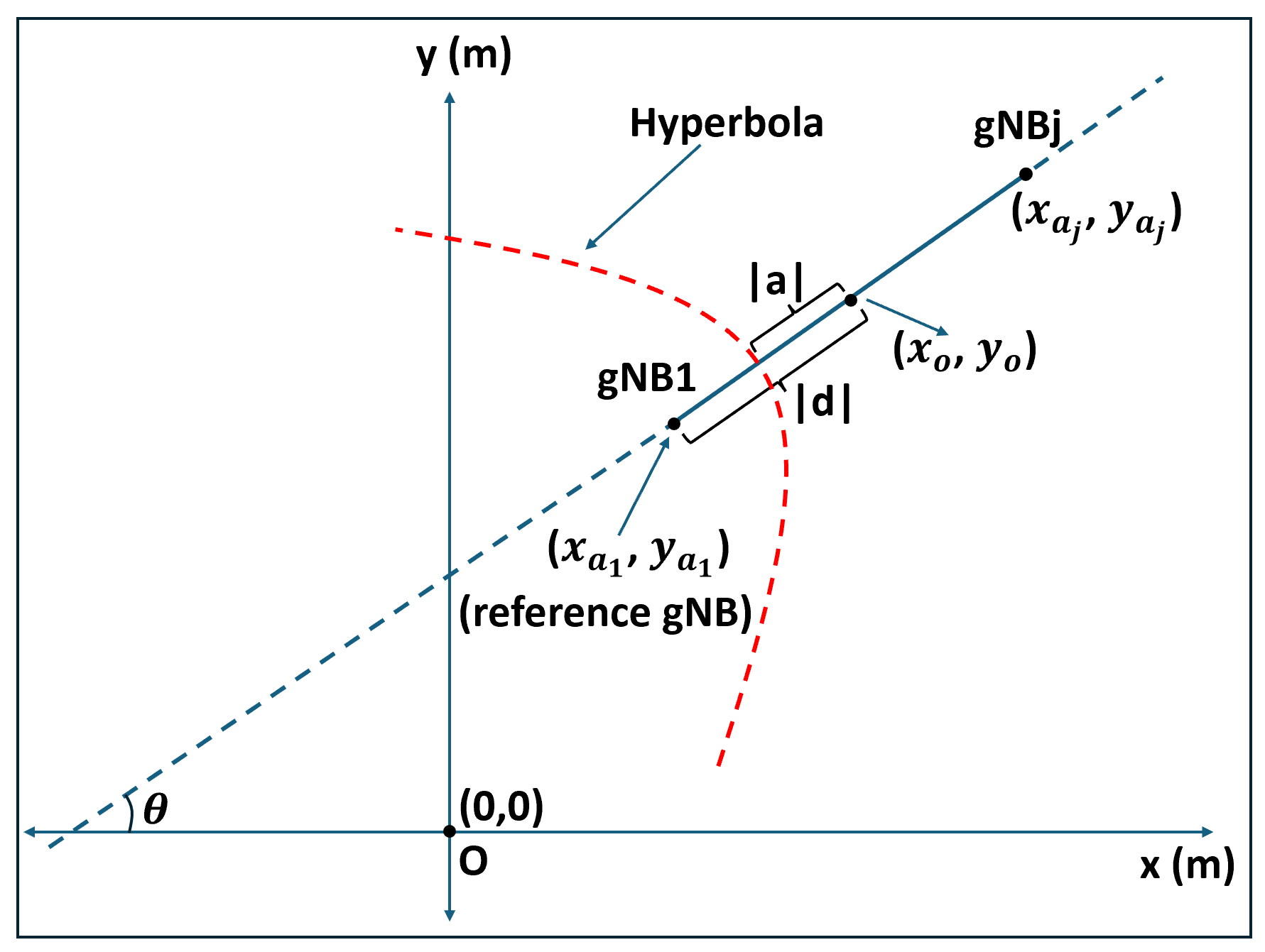}
  \captionsetup{font=small}
  \vspace{-0.1in}
  \caption{ Hyperbola formulation using two focii located at reference gNB $1$ and non-reference gNB $j$. 
  }
  \Description{hyperbola equation}
  \label{fig_hyperbola}
  \vspace{-0.01in}
\end{figure}

This section aims to estimate the UE position from the TOA measurements obtained from three gNBs at the UE. We assume that we have valid estimates of the inter-gNB time offsets $\hat{\Delta}_2$ and $\hat{\Delta}_3$, obtained via the calibration procedure explained in Section \ref{section_calibration}. We begin by showing the geometrical significance of the RSTD measurements. Let the unknown UE position be $\bm{\alpha}_i=[x_{n_i},y_{n_i}]^T$ and we have TOA measurements from reference gNB $1$ and gNB $j$ as $\tau_{i,1}$ and $\tau_{i,j}$ respectively. From the TOA measurements, we first obtain the measured RSTD followed by the corrected RSTD for every non-reference gNB $j$ as
\begin{align}
\label{eq_RSTD_from_toa}
RSTD_{i,j} = \tau_{i,j} - \tau_{i,1}, \;\; RSTD_{i,j}^\prime = RSTD_{i,j} - \hat{\Delta}_j, \; \forall j \ne 1. 
\end{align}
In other words, the measured RSTD values at the UE are erroneous and need to be corrected using the estimated inter-gNB time offsets. By subtracting the estimated inter-gNB offsets from the measured RSTD values, we obtain corrected RSTD values. In Figure \ref{fig_hyperbola}, we have the reference gNB $1$ positioned at $[x_{a_1},y_{a_1}]^T$ and the non-reference gNB $j$ positioned at $[x_{a_j},y_{a_j}]^T$. The UE lies on a hyperbola determined by the two foci lying on gNB $1$ and gNB $j$ with length of the semi-major and semi-minor axes as $|a|$ and $|b|$, respectively. Assuming the variables of the hyperbola are $x_{n_i}$ and $y_{n_i}$, the parametric equation of the branch of the hyperbola that opens towards the reference gNB $j=1$ is 
{\small
\begin{equation}
    \label{eq_hyperbola}
    \begin{split}
    &\begin{bmatrix}
    x_{n_i}(t) \\
    y_{n_i}(t)
    \end{bmatrix} 
    =
    \begin{bmatrix}
    \cos{\theta}&-\sin{\theta}\\
    \sin{\theta}&\cos{\theta}\\
    \end{bmatrix} 
    \begin{bmatrix}
    -|a|\cosh{t}\\
    b\sinh{t}\\
    \end{bmatrix} + 
    \begin{bmatrix}
    x_0\\
    y_0
    \end{bmatrix}, \\
    & a=\frac{c}{2}RSTD_{i,j}^\prime,\;\; b^2 = d^2-a^2, x_0=\frac{x_{a_j}+x_{a_1}}{2}, y_0=\frac{y_{a_j}+y_{a_1}}{2} \\
    &\theta = \arctan{\left(\frac{y_{a_j}-y_{a_1}}{x_{a_j}-x_{a_1}}\right)}, \;\; d=\frac{\sqrt{(x_{a_j}-x_{a_1})^2+(y_{a_j}-y_{a_1})^2}}{2}. 
    \end{split}
\end{equation}
}
Here, $t$ is the parameter, $\cosh{t}$ and $\sinh{t}$ are the hyperbolic cosine and sine functions. Note here, $d^2>a^2$ so that $b^2>0$. Ideally, if we obtain $a$ from the true RSTD value, it will give us the true hyperbola. However, since the true RSTD depends on the UE position which itself is unknown, we use the corrected RSTD measurements. Having valid estimates of the inter-gNB time offsets is crucial to obtaining a hyperbola close to the true hyperbola. Discussion about the validity of the inter-gNB time offset estimates is postponed to Section \ref{section_experimental_setup}. If for some reason we had the invalid estimates of the inter-gNB time offsets, the corrected RSTD values could be skewed such that in equation\eqref{eq_hyperbola}, $a^2>d^2$ and $b^2<0$ leading to an invalid hyperbola. This is a sufficient but not necessary condition for checking for the validity of the inter-gNB time offsets. 
\par
Therefore, for every non-reference gNB $j$ ($j \ne 1$), we obtain a hyperbola. The UE position is determined by finding the intersection point of the hyperbolas. Each hyperbola is determined by the relative distance of the gNB from the reference gNB and the corrected RSTD measurements at the UE. Even if valid estimates of the inter-gNB time offsets are used, the corrected RSTD measurements may still not match the true RSTD values. This discrepancy arises due to residual errors in estimating the inter-gNB time offsets and noise in the TOA measurements. Consequently, the hyperbolas may not intersect and we could go with a least squares approach. We used the function $GetEstimatedUEPosition$
 in the {\em Matlab 5G Toolbox} to obtain the intersection point of the two hyperbolas.

\section{Experimental Results and Discussion}
\subsection{Experimental Procedure}
\label{section_experimental_setup}
For each device, there is an SDR which is connected to the octoclock for time and frequency synchronization. This synchronization happens as soon as the OAI software stack is initialized on the device computer. Hence, as soon as the gNBs are initialized, $\Delta_2$ and $\Delta_3$ are unknown and need to be obtained using the calibration procedure explained in Section \ref{section_calibration}. The UE can be initialized as and when required since the UE offset $\phi$ is not required to be known.
\par
In OAI, we encountered difficulties in obtaining valid channel estimates from three gNBs while they were simultaneously transmitting. We suspect this is due to inter-gNB interference. To combat this, for experiments requiring multiple gNBs, we made the following modifications to the OAI software stack. We scheduled the PRS transmissions from each gNB to transmit one at a time for a period of five seconds. Thus at any given time we have one active gNB transmitting and valid channel estimates for the transmitting gNB can be obtained at the UE. Note, this has no bearing on the time offsets since the transmissions always always occur at a frame boundary and the TOA measurements correpond to the signal propagation delay plus the random static time offset introduced by the octoclock.

\subsection{Multipath Effects}
In this section, we investigate the impact of multipath on TOA estimation using channel estimates for a one gNB - one UE test setup. 
There are two different scenarios in which we conducted experiments. In the first scenario, the testbed was setup in an outdoor environment -- on a rooftop to minimize mutipath signal propagation. In the second scenario, the testbed was setup in an indoor environment which would experience significant multipath signal propagation.

\begin{figure*}[t!]
     \centering
     \begin{subfigure}[b]{0.40\linewidth}
         \centering
         \includegraphics[clip, trim=0.3cm 0.1cm 0.2cm 0.2cm, width=\textwidth]{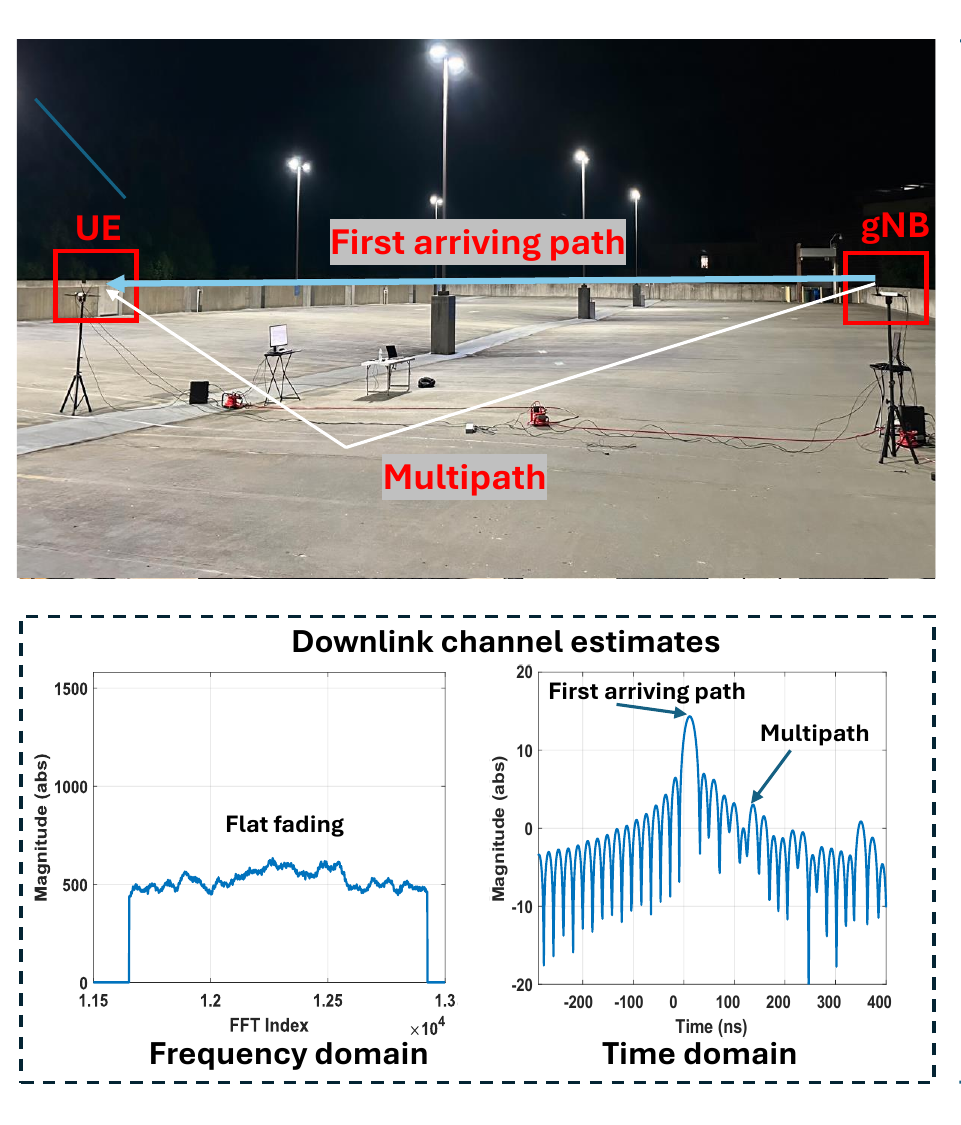}
         \caption{\footnotesize Minimal-multipath scenario (Outdoors)}
         \label{fig_minimal_mutipath}
     \end{subfigure}
     \begin{subfigure}[b]{0.40\textwidth}
         \centering
        \includegraphics[clip, trim=0.3cm 0.1cm 0.2cm 0.2cm,  width=\textwidth]{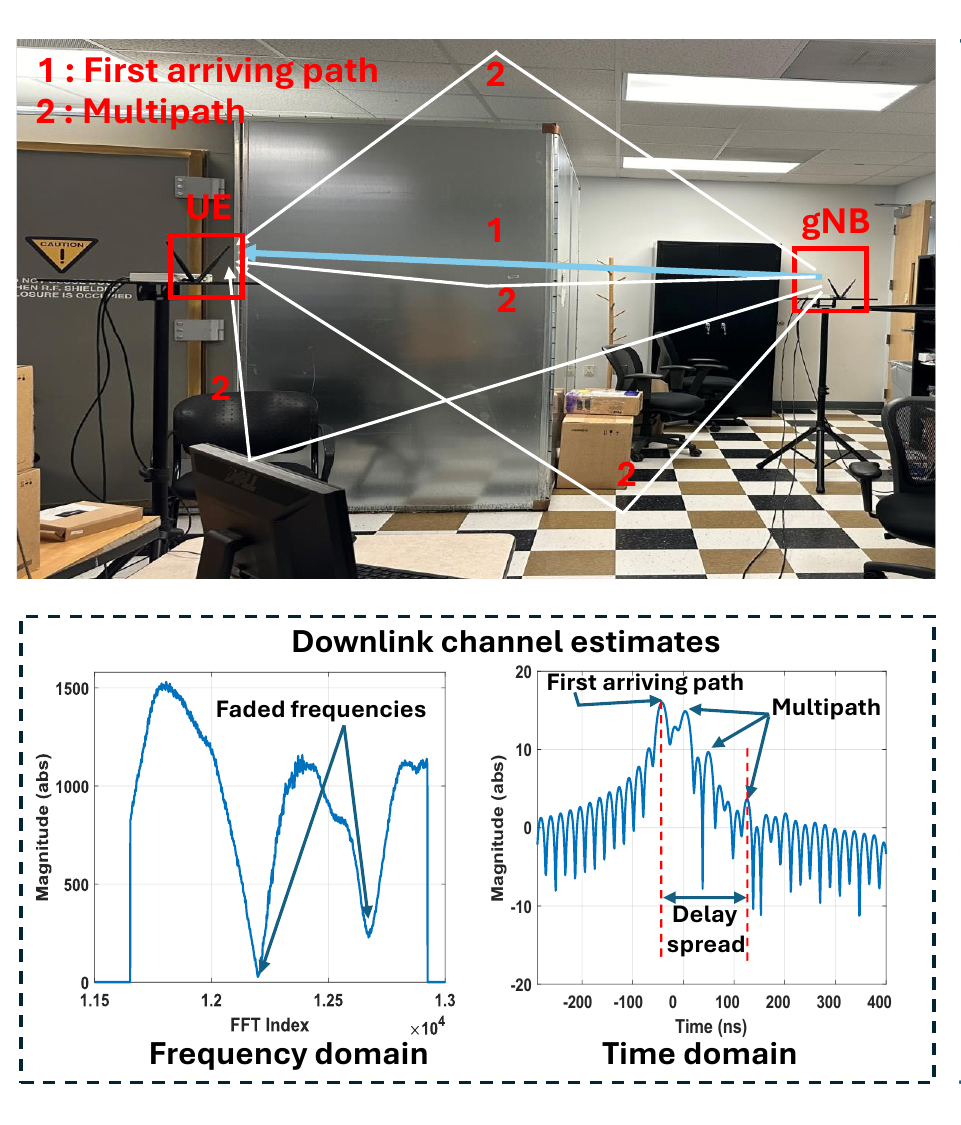}
        \caption{\footnotesize Significant-multipath scenario (Indoors)}
        \label{fig_multipath}
     \end{subfigure}
     \captionsetup{font=small}
     \vspace{-0.1in}
     \caption{Effect of multipath on TOA estimation from the downlink channel estimate for (a) Minimal-multipath scenario and (b) Significant-multipath scenario.}
     \label{fig_multipath_effects}
     \Description{Channel response}
     \vspace{-0.1in}
\end{figure*}

 In Figure \ref{fig_multipath_effects}, we show the frequency and time domain downlink channel estimates at the UE in both minimal-multipath and significant-multipath environments. Observe, in Figure \ref{fig_minimal_mutipath} representing the minimal-multipath scenario, the frequency domain channel estimate is relatively flat. Further, the time domain channel representation shows a single peak corresponding to the direct path. {\em Digital oversampling} in this case effectively sharpens this peak, improving the TOA estimation accuracy.
 \par
 In contrast, in Figure \ref{fig_multipath} which represents the significant -multipath case, the received signal at the UE is a superposition of multiple copies of the transmitted PRS signal, each arriving at different UE times due to multiple signal propagation paths. This manifests as multiple closely spaced overlaping peaks in the time domain channel estimate. Each peak corresponds to a different propagation path formed by a combination of reflections, diffractions, and scattering from different environmental obstacles \cite{Gifford2020}. The spacing between these peaks depends on the geometrical arrangement of the environmental obstacles and can be characterized for different environments using {\em delay spread} as a metric. This effect is called {\em multipath fading} since the frequency domain channel representation exhibits `fading' or loss of certain frequency components as shown in Figure \ref{fig_multipath}. 
 
 \par
To tackle this challenge, we need to increase the PRS signal bandwidth, which reduces peak width and overlap, thus improving TOA resolution. Alternatively, super-resolution algorithms \cite{li2004super} can help mitigate the adverse effects of multipath on TOA estimation. In summary, while digital interpolation enhances TOA accuracy in minimal-multipath scenarios, significant multipath presents substantial challenges. Addressing these challenges requires characterizing the delay spread in different environments to appropriately select signal bandwidth.
\vspace{-0.1in}
\subsection{Characterization of UE-gNB/Inter-gNB Timing Offsets}
\label{section_results_time_offsets}
\par
From each channel estimate, we can extract a TOA estimate. Currently, we obtain one channel estimate per gNB for every PRS Resource set period, which equates to every 20 slots or $20 \times 0.5 = 10$ ms. According to our modified OAI code, each gNB transmits PRS sequentially, one at a time, for five seconds each. Thus, we obtain $\frac{5}{10^{-2}}=500$ channel estimates per gNB. To estimate the inter-gNB timing offsets, we place the gNBs at known locations and then we initialize all of them. Since the time and frequency synchronization happens only when the 5G software stack is initialized, we expect all the inter-GnB offsets should remain static over time. Due to this reason, after the gNBs are started they are left untouched and at every moment, we have exactly one gNB transmitting PRS. The UE is aware of the gNB that is transmitting. Once we obtain 500 channel estimates for each gNB over the course of $15$ seconds, we declare a trial as completed and the UE is turned off. Note, since the 5G software on the UE device is reinitialized for every trial, the UE-gNB time offset $\phi$ changes across trials, whereas the inter-gNB offset remains constant in agreement with the 3GPP specifications. 

\begin{figure}[t]
     \centering
     \begin{subfigure}[b]{0.49\linewidth}
         \centering
         \includegraphics[clip, trim=0.3cm 0.1cm 0.2cm 0.2cm, width=\textwidth]{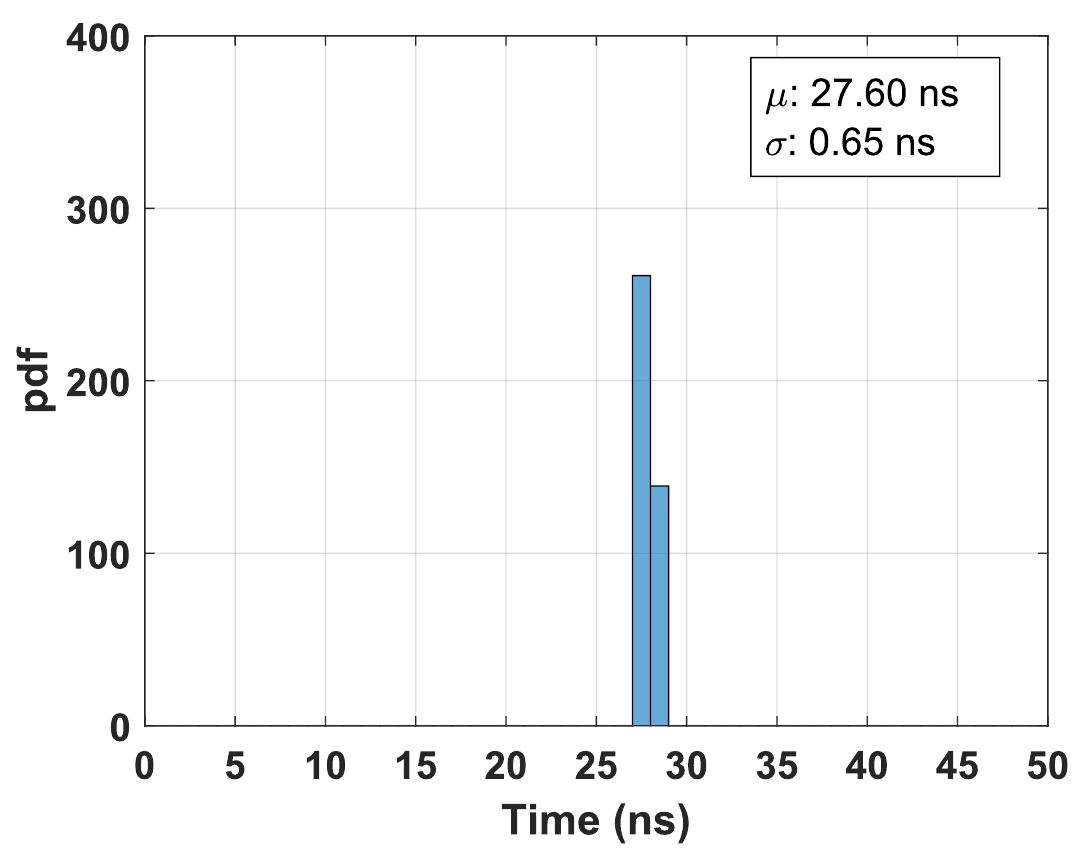}
         \caption{\footnotesize Trial 1 Histogram}
         \label{fig_histogram_toa_trial1}
     \end{subfigure}
     \hfill
     \begin{subfigure}[b]{0.49\linewidth}
         \centering
        \includegraphics[clip, trim=0.3cm 0.1cm 0.2cm 0.2cm,  width=\textwidth]{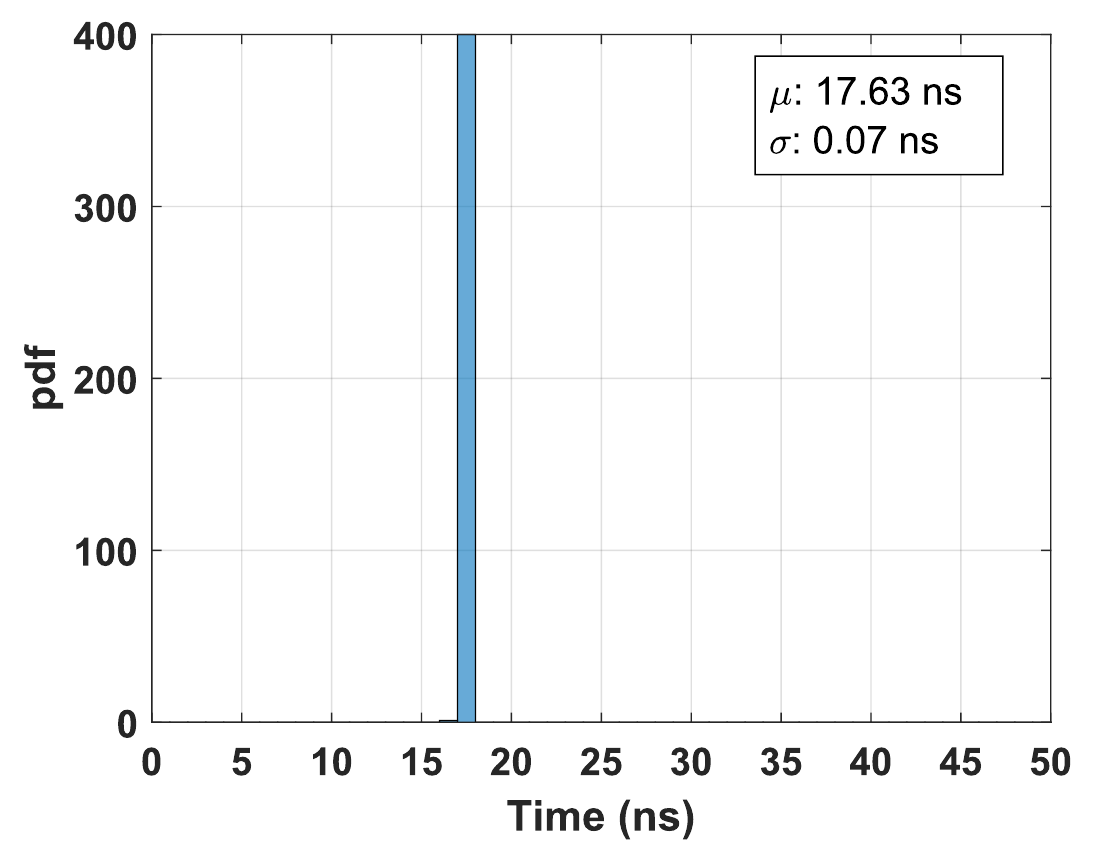}
        \caption{\footnotesize Trial 2 Histogram}
        \label{fig_histogram_toa_trial2}
     \end{subfigure}
     \hfill
     \captionsetup{font=small}
     \vspace{-0.2in}
     \caption{Histogram constructed using 500 TOA values. 
     }
     \label{fig_histogram_toa}
     \Description{TOA histogram}
 \end{figure}
 \par
 We conducted two trials for the same UE location and for each trial we produced a histogram of the TOA values across 500 channel estimates obtained from a fixed gNB. The histogram is shown in Figure \ref{fig_histogram_toa}. Observe, the mean value changes from $27 $ ns in trial $1$ to $17$ ns in trial $2$ for the same UE position. This is because the UE has been reinitialized between the trials and the TOA values have an arbitrary UE-gNB offset $\phi$ leading to the shift in the histogram. Now within a trial, note that the standard deviation of the histogram is very small - 0.65ns in Trial $1$ and 0.07 in Trial $2$. This means that over the course of a trial, both the UE-gNB offset and the inter-gNB offset remain constant. 
 
 Moving onto the measurement campaign to obtain estimates $\hat{\Delta}_2,\hat{\Delta}_3$. We placed the three gNBs at fixed known locations and initialized them. Now, the UE was placed at $K=9$ known locations $\alpha_i,\;\; i \in [1,K]$. At every UE location a trial was conducted i.e. the UE initialized and 500 channel estimates were collected per gNB. From these TOA estimates are extracted and used to estimate the inter-gNB time offsets according to Section \ref{section_calibration}. The estimated values are $\hat{\Delta}_2 = 41.2\;\text{ns}$ and $\hat{\Delta}_3 = 30.9\;\text{ns}$. These timing offsets are within the $\pm50$ ns accuracy provided by the octoclock specification. Note, by using equation \eqref{eq_hyperbola}, these timing offsets lead to an error of $6.18$m and $4.65$m in the length of the semi-major and semi-minor axes of their corresponding hyperbolas which is significant.

\subsection{Positioning Results}
\begin{figure}[b]
\centering     
\begin{subfigure}[b]{0.49\linewidth}
         \centering
         \includegraphics[width=\textwidth]{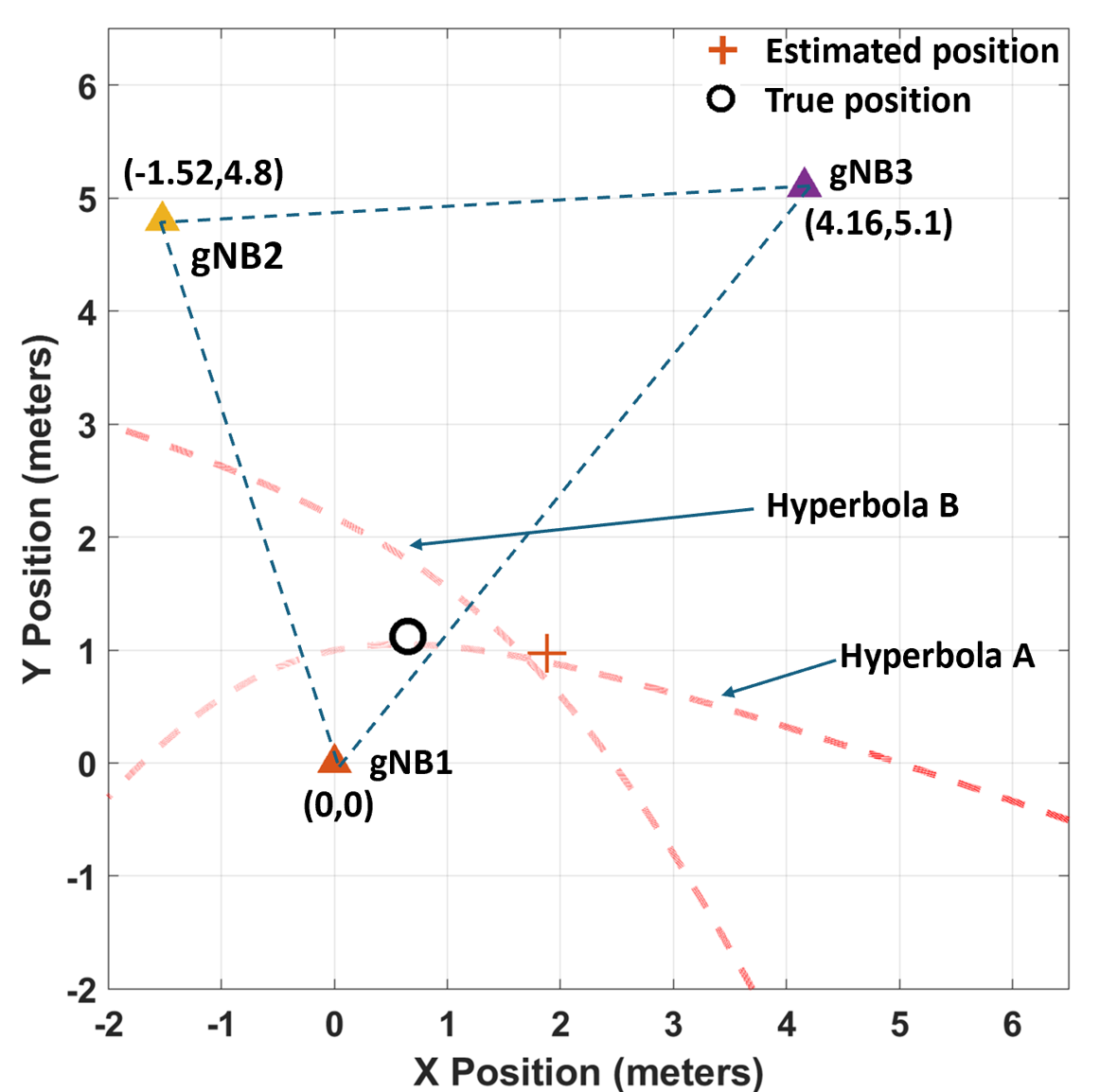}
         \caption{\footnotesize UE position estimation}
         \label{fig_positioning}
\end{subfigure}
     \hfill
     \begin{subfigure}[b]{0.49\linewidth}
         \centering
         \includegraphics[width=\textwidth]{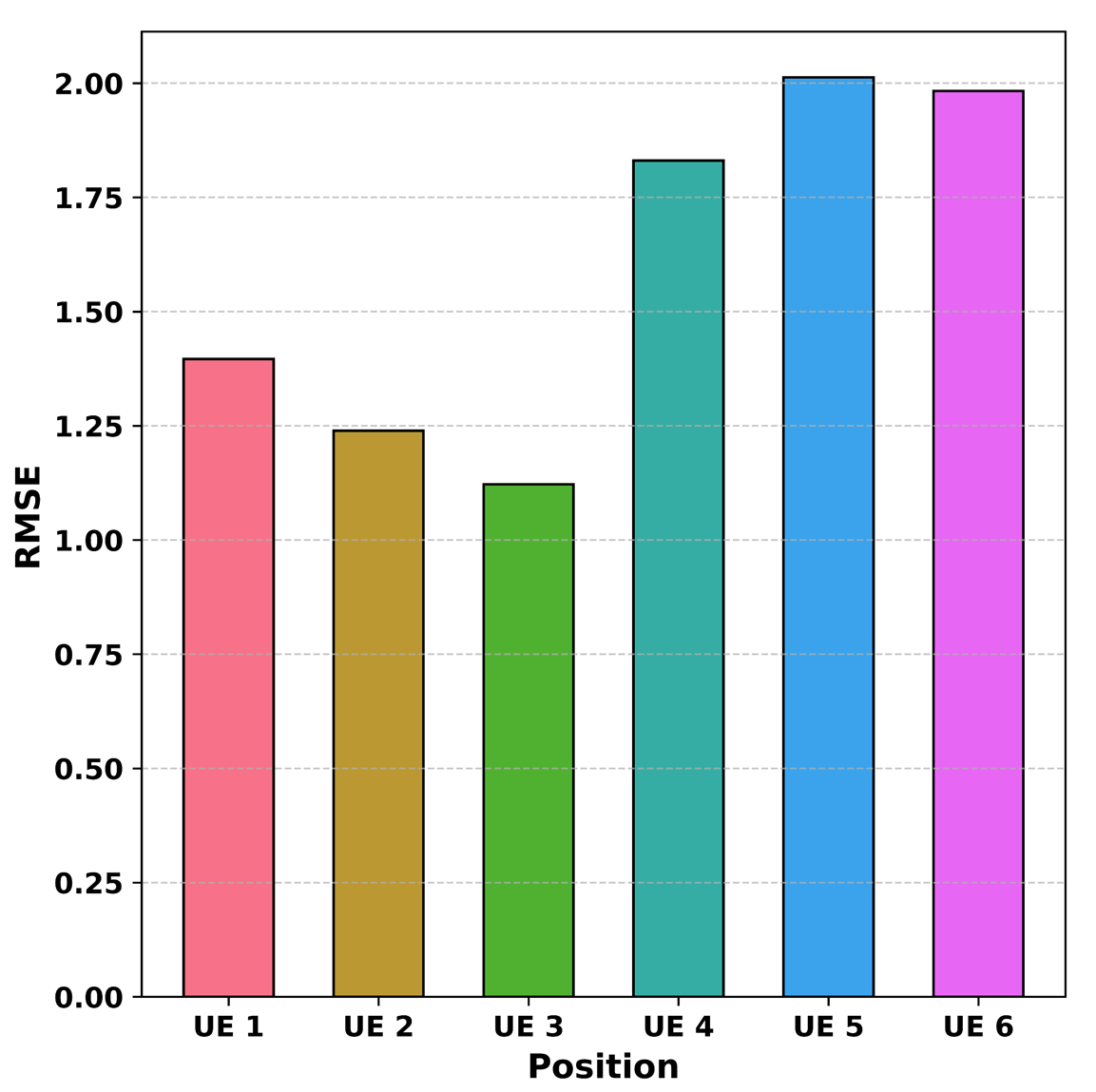}
         \caption{\footnotesize RMSE plot}
         \label{fig_RMSE}
\end{subfigure}
\captionsetup{font=small}
\vspace{-0.1in}
\caption{\footnotesize UE position estimation using the method described in section \ref{section_tdoa_positioning} and mean RMSE error of $1.6$m.}
         \label{fig_positioning_results}
         \Description{RMSE}
\vspace{-0.1in}
         
\end{figure}
To achieve positioning, we first employ the measurement campaign outlined in section \ref{section_results_time_offsets} to obtain estimates of the inter-gNB timing offsets. The validity of the inter-gnB offsets is maintained as long as the gNBs remain operational after obtaining $\hat{\Delta}_2$ and $\hat{\Delta}_3$ otherwise this step needs to be repeated. Now, the UE is placed at an arbitrary position $\bm{\alpha}_i=[x_n,y_n]^T$ and we acquire RSTD measurements from the two non-reference gNBs i.e. $RSTD_{i,j}, \; j\in \{2,3\}$ from fresh TOA measurements according to equation \eqref{eq_RSTD_from_toa}. Next, using the valid estimates of the inter-gNB timing offsets, we obtain corrected RSTD measurements from the same non-reference gNBs according to equation \eqref{eq_RSTD_from_toa}. From the known gNB positions and the corrected RSTD measurements, we can obtain two hyperbolas - hyperbola A and hyperbola B shown in Figure \ref{fig_positioning}. The UE position can be estimated by finding out the intersection of the two hyperbolas. The Root-Mean-Squared-Error (RMSE) between $6$ estimated UE positions to the true UE positions is shown in Figure \ref{fig_RMSE}.   
\vspace{-0.1in}
\section{Conclusion }
In this paper, we developed a 3GPP-compliant 5G positioning testbed and analyzed a TOA measurement model that includes (a) inter-gNB offsets and (b) UE-gNB time offsets. We examined the impact of inter-gNB offsets and assessed their validity and stability. Additionally, we experimentally studied the influence of multipath on TOA estimation, showing that digital interpolation improves accuracy in low-multipath scenarios but is ineffective in high-multipath environments. Using a calibration method for estimating inter-gNB offsets, we achieved 2D positioning with an RMSE of 1.6 m in the tested scenarios.
\vspace{-0.1in}

\section{Acknowledgement}
This work was prepared by the authors using Federal funds under award \#70NANB22H070 from the National Institute of Standards and Technology (NIST) Public Safety Communications Research (PSCR) Division’s Public Safety Innovation Accelerator Program (PSIAP), U.S. Department of Commerce and the NIJ graduate research fellowship through grant 15PNIJ-23-GG-01949-RES. The statements, findings, conclusions, and recommendations are those of the author(s) and do not necessarily reflect the views of the NIST PSIAP PSCR or the U.S. Department of Commerce. 
\bibliographystyle{ACM-Reference-Format}
\footnotesize
\bibliography{sample-base}

\appendix
\end{document}